\documentclass[runningheads]{llncs}

\usepackage[T1]{fontenc}

\usepackage[usenames,dvipsnames]{xcolor}
\usepackage{cite}

\usepackage{hyperref}

\usepackage[utf8]{inputenc}%
\usepackage{xspace}
\usepackage{cleveref}
\usepackage{color}
\usepackage{url}
\usepackage{adjustbox}
\usepackage{booktabs}
\usepackage{wrapfig}

\usepackage[absolute,showboxes]{textpos}

\newcommand{\df}[1]{\textcolor{blue}{#1}}
\renewcommand{\df}[1]{#1}

\usepackage[absolute,showboxes]{textpos}

\newcommand*{\eg}{\emph{e.g.,}\@\xspace}
\newcommand*{\Eg}{\emph{E.g.,}\@\xspace}
\newcommand*{\etal}{\emph{et al.}\@\xspace}
\newcommand*{\ie}{\emph{i.e.,}\@\xspace}
\newcommand*{\viz}{\emph{viz.}\@\xspace}

\newcommand{\p}[1]{\noindent\textbf{#1}\xspace}

\newcommand*{\tool}{\emph{BannerClick}\@\xspace}
\newcommand*{\openwpm}{OpenWPM\@\xspace}

\newcommand{\one}{(1)~}
\newcommand{\two}{(2)~}
\newcommand{\three}{(3)~}
\newcommand{\four}{(4)~}

\begin{document}
\setlength{\TPHorizModule}{\paperwidth}
\setlength{\TPVertModule}{\paperheight}
\TPMargin{5pt}
\begin{textblock}{0.8}(0.1,0.02)
     \noindent
     \footnotesize
     If you cite this paper, please use the PAM reference:
     Ali Rasaii, Shivani Singh, Devashish Gosain, and Oliver Gasser.
     Exploring the Cookieverse: A Multi-Perspective Analysis of Web Cookies.
     In \textit{Proceedings of the Passive and Active Measurement Conference (PAM ’23), March 21--23, 2023}.
\end{textblock} 
 
\title{Exploring the Cookieverse:\\ A Multi-Perspective Analysis of Web Cookies}

\author{Ali Rasaii\inst{1} \and
Shivani Singh\inst{2} \and
Devashish Gosain\inst{1,3} \and
Oliver Gasser\inst{1}}
\institute{Max Planck Institute for Informatics \and
New York University \and
KU Leuven \\
\email{\{arasaii,oliver.gasser\}@mpi-inf.mpg.de \\
shivani.singh@nyu.edu \\
dgosain@esat.kuleuven.be
}}

\maketitle

\begin{abstract}

Web cookies have been the subject of many research studies over the last few years. However, most existing research does not consider multiple crucial perspectives that can influence the cookie landscape, such as the client's location, the impact of cookie banner interaction, and from which operating system a website is being visited. In this paper, we conduct a comprehensive measurement study to analyze the cookie landscape for Tranco top-$10$k websites from different geographic locations and analyze multiple different perspectives.
One important factor which influences cookies is the use of cookie banners.
We develop a tool, \tool, to automatically detect, \df{accept, and reject cookie banners with an accuracy of 99\%, 97\%, and 87\%, respectively}.
We find banners to be 56\% more prevalent when visiting websites from within the EU region.
Moreover, we analyze the effect of banner interaction on different types of cookies (\ie first-party, third-party, and tracking). For instance, we observe that websites send, on average, $5.5\times$ more third-party cookies after clicking ``accept'', underlining that it is critical to interact with banners when performing Web measurements.
Additionally, we analyze statistical consistency, evaluate the widespread deployment of consent management platforms, compare landing to inner pages, and assess the impact of visiting a website on a desktop compared to a mobile phone.
Our study highlights that all of these factors substantially impact the cookie landscape, and thus a multi-perspective approach should be taken when performing Web measurement studies.

\end{abstract}

\section{Introduction}

Web cookies serve various purposes, like keeping the user logged in or storing a user's website settings. However, other than their originally intended use, cookies have been exploited for commercial activities like user tracking and advertisement targeting~\cite{acar2014web, englehardt2015cookies, englehardt2016online, bangera2017ads, razaghpanah2018apps}.
As a consequence, various data protection laws have been enacted in the past few years, \eg the General Data Protection Regulation (GDPR) \cite{GDPR} in the EU and the California Consumer Privacy Act (CCPA) \cite{CCPA} to regulate the use of cookies.

Numerous studies shed light on the complex ecosystem of sharing users’ personal information across various third parties \cite{li2015trackadvisor,schelter2016tracking, Lerner2016, cahn2016empirical, gonzalez2017cookie} and to what extent GDPR mitigates such abuse \cite{trevisan2019}.
However, most of this research was conducted from a single or a limited number of vantage points (VPs). %
Thus, in this work, we characterize the cookie landscape from diverse geographic locations spanning six continents---North America, South America, Europe, Africa, Asia, and Australia. 
We complement the existing research by globally analyzing the following aspects of the cookie landscape:

\noindent \textbf{Interaction with cookie banners:} Most research involving GDPR does not consider interaction with cookie banners (\eg clicking accept/reject buttons) \cite{acar2014web, englehardt2015cookies, trevisan2019,linden2020privacy}. Thus, we develop the automated tool \tool to automatically \df{detect, accept and reject cookie banners with an accuracy of 99\%, 97\%, and 87\%, respectively} (see \Cref{sec:approach}).
With \tool we automatically detect banners on about $47\%$ of the Tranco top-$10$k websites in the EU region whereas in non-EU regions we find banners on less than $30\%$ of websites (see \Cref{sec:effectsbanners}).
Furthermore, we analyze the difference in the number of cookies before and after interacting with a cookie banner and find an increase of 5.5$\times$ for third-party cookies.

\noindent \textbf{Impact of geographic locations:} To assess the effectiveness of GDPR, we compare observed cookies (especially third-party and tracking cookies) between EU and non-EU vantage points (cf. \Cref{sec:location}). 
We find that without banner interaction, 43\% of websites send more tracking cookies when accessed from non-EU regions compared to the EU.
Even after accepting a banner, 83\% of websites send more tracking cookies in non-EU countries compared to EU countries.
This percentage increases to 96\% when rejecting banners.
Our findings indicate a positive impact of GDPR on reducing the number of TP and tracking cookies.

\noindent \textbf{Consistency of websites:} 
For cookie analysis, it is essential to observe that when a website is accessed multiple times, it sets a consistent number of cookies. If the variation in the number of cookies is high, then one cannot have statistically significant deductions about cookie characteristics (\eg number of third-party cookies).
Thus we perform two statistical tests: First, we use the coefficient of variation to test for intra-location consistency, \ie how consistent the cookie landscape is when visiting a website multiple times from the same location.
Second, we use the Mann–Whitney U test \cite{mann1947test} to test for inter-location consistency, \ie how consistent is the cookie landscape when visiting a website from different locations.
Our results show that websites are more consistent within the EU and that we find the most statistically significant differences between EU and non-EU countries (cf. \Cref{sec:consistency}).

\noindent \textbf{Cookie differences between landing and inner pages:} We also explore the difference in cookies between the landing and inner pages of a website (see \Cref{sec:landVsInner}). As shown by previous work, the structure and content of landing pages differ substantially from inner pages~\cite{aqeel2020landing}.
Similarly, some websites may not send cookies on landing pages but may send them on inner pages. %
Hence, we quantify the difference between cookies on the landing and the inner pages of a website.
For instance, at our United States VP, we find that 32\% of websites send more third-party cookies on the landing compared to inner pages.
Similarly, 29.7\% of websites send more third-party cookies on inner pages when accessed from Germany.
Overall, we find that 27.4\% and 15.7\% of websites exhibit different third-party and tracking cookie behavior on all VPs.
Thus, studies analyzing \textit{only} the landing pages may not present the full picture of the cookie landscape.

\noindent \textbf{Cookie differences when a website is accessed from desktop and mobile browsers:}
As mobile Web browsing is becoming more popular and overtaking desktop browsing \cite{mobile-desktop,mobile-desktop-statista}, it is important to study its cookie differences.
This is underlined by the fact that websites often have mobile-specific versions that could lead to a difference in cookies.
Thus, we conduct measurements to quantify the cookie differences between mobile and desktop (cf. \Cref{sec:mobilevsdesktop}).
For instance, our US East VP sees more third-party cookies on desktop compared to mobile for 28\% of all websites.
Contrarily, when accessing websites from Brazil, 28\% set more third-party cookies on mobile.
Overall, 14.6\% and 9\% of websites show different third-party and tracking cookie behavior on all VPs.
Therefore, future research investigating cookie behavior needs to take desktop as well as mobile websites into account.

Additionally, we analyze the \textit{impact of the Brazilian and Californian privacy laws} \cite{CCPA, schreiber2020right} on Web cookies.
Since these laws came into effect recently (\ie in 2020), the analysis of their impact is still in its early days \cite{Chen2021,Connor2021}.
Following California's privacy law, other US states are also considering adopting online privacy laws \cite{van2022setting}.
Thus it becomes necessary to draw insights from the enactment of these existing laws on the cookie landscape.
In \Cref{sec:CCPA}, we show that CCPA does not have a direct positive impact on Web cookies.
Instead, we find that websites publicly adhering to CCPA tend to send more third-party and tracking cookies compared to others.

Overall, our measurement study highlights that factors like banner interaction, client location, landing vs. inner pages, and desktop vs. mobile substantially impact Web cookies.
Thus, future research should incorporate these factors when analyzing the cookie landscape.
To encourage reproducibility, \df{we open-source our code \cite{bannerclick} and release our data and analysis scripts \cite{edmond} at \href{https://bannerclick.github.io}{\textbf{\texttt{bannerclick.github.io}}}.}

\section{Background}

In this section, we provide background information on different privacy laws and Web measurement platforms.

\subsection{Privacy Laws Regarding Web tracking}

\noindent\textbf{General Data Protection Regulation (GDPR):}
The European Union's GDPR---which came into effect in May 2018---is considered to be one of the most comprehensive laws safeguarding user privacy online.

The GDPR mandates that the storage and exchange of personal information (\eg cookies)
is allowed only after a user has explicitly consented. The only exception is for ``strictly necessary'' cookies that are essential for a website's operation, \eg storing user credentials. According to the GDPR, websites must obtain users’ consent concisely and transparently. This results in websites showing \textit{cookie banners}, informing users about the cookies being collected by the websites and third parties. Some banners explicitly ask for users' consent (\eg with \textit{accept} or \textit{reject} buttons), and some assume users' continued website use as implied consent.
In this research, we study the impact of GDPR on cookie characteristics across the globe. 

\noindent\textbf{California Consumer Privacy Act (CCPA):}
CCPA is a state statute enacted by the California state assembly in June 2018. CCPA has similar goals as GDPR: it intends to protect the privacy of the residents of California. CCPA enables California residents to know what personal data is being collected (\eg their IP address), whether it is being sold to third parties, and the right to refuse to share their data.
All companies operating in California with at least an annual revenue of \$25 million must comply with the law. Importantly, even if these companies are not headquartered in California (or even the US), they still come under the purview of the CCPA.

\noindent\textbf{Brazil's General Personal Data Protection Law (LGPD):}
Similar to the EU, Brazil also introduced a privacy law ``Lei Geral de Proteção de Dados Pessoais'' (LGPD) \cite{schreiber2020right,LGPD-diff} that was enforced on September 2020. LGPD is again similar to GDPR. It also focuses on personal data and users' rights. Moreover, it states that website publishers must obtain consent before storing the personal data of clients (in the form of cookie banners). %

To the best of our knowledge, in this work, we take the first step to empirically quantify the impact of CCPA and LGPD on Web cookies.

\subsection{Web privacy measurement platforms}
There exist a variety of Web privacy measurement platforms, \eg \openwpm \cite{englehardt2016online}, FPdetective \cite{acar2013fpdetective}, Chameleon \cite{Chameleon} and Common Crawl \cite{CommonCrawl}.

\openwpm is built using Python and uses the Firefox Web browser to visit websites through the Selenium automation tool \cite{selenium}. 
\openwpm is feature-rich, provides speed and scalability for large-scale measurements \cite{englehardt2016online}, and has been used by a plethora of Web measurement studies \cite{OpenWPM_popular}.
Thus in our research, we use and extend \openwpm to collect, store, and analyze measurement data.

\section{Data Collection and Approach}
\label{sec:approach}

We now present our VP locations, target websites, and our approach to studying the cookie landscape in detail.

\subsection{Location Diversity and Target Websites}
\label{subsec:locations}
We use AWS cloud instances at the following locations as our VPs: Frankfurt (Germany), Stockholm (Sweden), Ashburn (US East), San Francisco (US West), Mumbai (India), São Paulo (Brazil), Cape Town (South Africa), and Sydney (Australia). We select these vantage points to have two VPs inside GDPR countries (Germany and Sweden), two VPs in the US (of which one is in the CCPA state California), one in Brazil (that has LGPD), one in Africa, one in Australia, and one in Asia. %

In our measurement study, we use the global Tranco top-$10$k \cite{Tranco} as target websites for our analysis. The popularity of these websites is measured considering the actual Web traffic of users \cite{scheitle2018long}. Other counterparts like the Cisco Umbrella list \cite{Umbrella} and the Majestic Million list \cite{MajesticMillion} are created using indirect sources like DNS queries and URLs embedded in website ads. %

Additionally, for some experiments that require repeated measurements (\eg consistency tests), we use a subset of Tranco top-$10$k websites; we select three sets of websites: Tranco top-$100$, $1001$--$1100$, and $9901$--$10$k. These sets include websites from the top, middle, and bottom of the Tranco top-10k websites and hence represent different website tiers. We call this subset the ``tiered Tranco list''.
In order to identify a suitable \openwpm configuration, we perform multiple small-scale test runs.
\Cref{tab:measurements} shows an overview of our final large-scale measurement runs.
The longest measurement takes 20 days, in which the Web can change substantially.
In order to keep results comparable, we ensure that each website is crawled at a similar time from all vantage points. \df{In the case of failure in one vantage point the website would be excluded from the final result.}
Moreover, we run \openwpm in stateless mode and ensure that the browser \df{does not block tracking} when accessing websites \cite{trackingblocking}.

As already mentioned, we completely automate our measurement campaign and access the Tranco websites using \openwpm. %
We now explain our approach to detecting and interacting with cookie banners on our target websites.

\begin{table}
\caption{Overview of different measurement types.}
\label{tab:measurements}
\centering
\begin{adjustbox}{max width=\columnwidth}
\begin{tabular}{llll}
\toprule
\textbf{Measurement Type} & \textbf{Start Date} & \textbf{Duration} & \textbf{Target Websites}   \\
\midrule
 Banner Interaction      &      Jan 20, 2022               &    20 days               &    Tranco Top 10k               \\
 Consistency Tests   &      Feb 9, 2022               &     10 days              &    Tranco tiered 300               \\%
 Landing vs. Inner    &     Mar 8, 2022                &    4 days               &   Tranco tiered 300                \\
 Desktop vs. Mobile   &     Feb 27, 2022                &       10 hours            &   Tranco tiered 300                \\%
 Impact of CCPA      &      Mar 13, 2022               &    10 hours               &    Tranco tiered 300               \\
\bottomrule
\end{tabular}
\end{adjustbox}
\end{table}

\subsection{Automated Banner Detection and Interaction}
\label{subsec:BannerClick}

Due to the EU ePrivacy Directive \cite{ePD} and GDPR \cite{GDPR}, many popular websites explicitly show cookie banners when accessed from within the EU \cite{Matte2020respect}. 
These banners must inform the user about what user data will be collected by the website (using cookies). Moreover, they must provide a clear choice to users on whether to accept or reject these cookies. 

To test whether websites respect the users' consent or not, (1) we detect the banners, (2) interact with them (\eg accepting/rejecting the banner policies), and (3) throughout the whole process collect all cookies.
We completely automate this process by developing our tool called ``\tool''. We now explain how our tool detects and interacts with cookie banners using Selenium browser automation \cite{selenium}.

To detect banners, we first create a corpus of English words that very likely exist in banners by manually inspecting $50$ random websites from Tranco top-$100$ domains. 
The corpus has eight unique English words \ie cookies, privacy, policy, consent, accept, agree, personalized, and legitimate interest. We translate these words into $11$ different languages (German,
Swedish,
Spanish, Italian, Portuguese, Chinese, Russian, Japanese, French, Turkish, and Persian) and append the translated words to the corpus, increasing the corpus size to $80$ words. We later show that with these words, we achieve an accuracy of about $99\%$ for detecting banners. 

\subsubsection*{\textbf{Banner Detection:}}
\label{subsec:banner_detection}

\begin{figure}[t!]
   \centering
   \includegraphics[width=0.7\columnwidth]{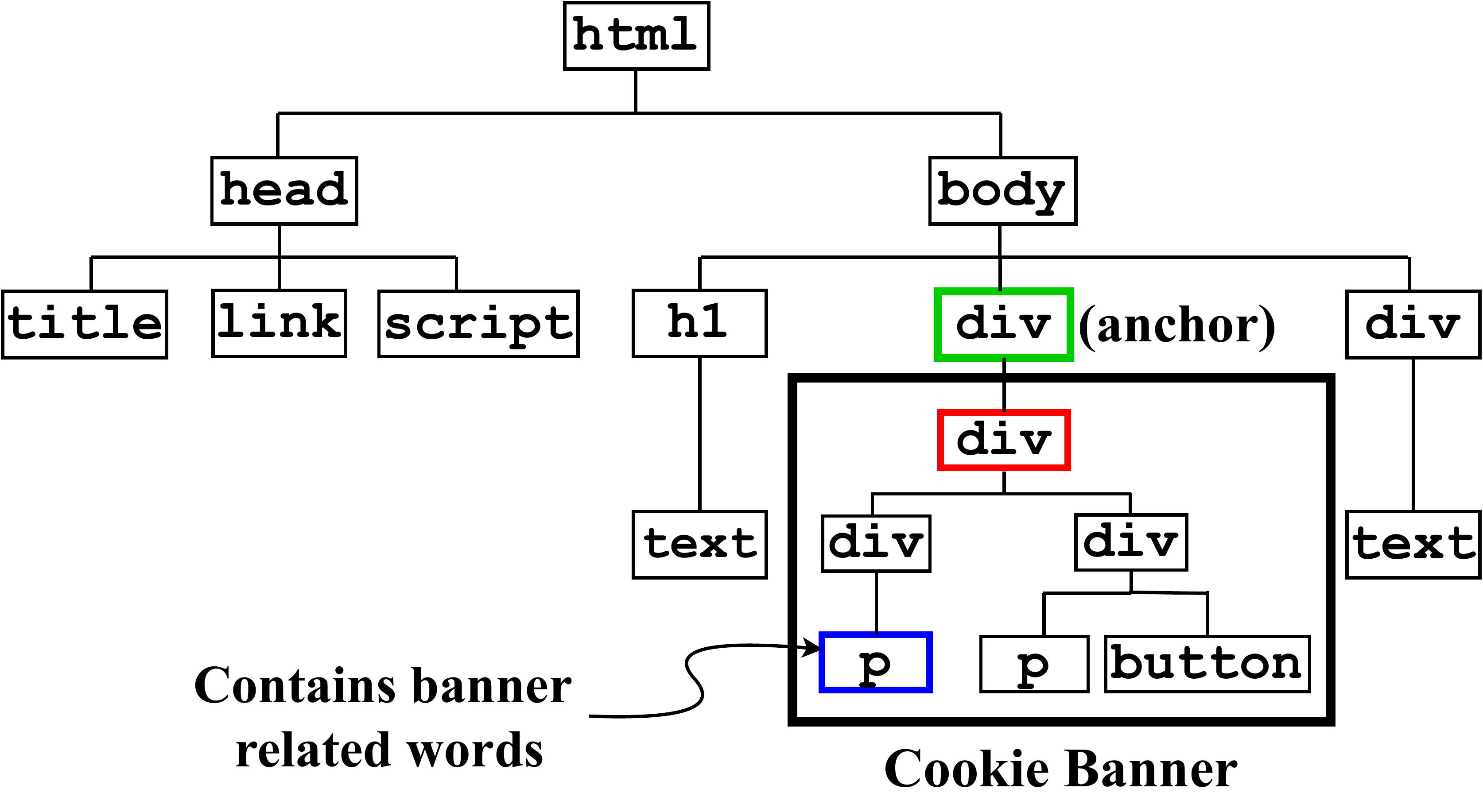}
    \caption{An HTML Document Object Model (DOM) containing a banner.}
    \label{fig:DOM}
\end{figure}

On a website's HTML page, \tool first searches all elements that contain a word from our corpus. As an example, the element \texttt{<p>} (shown in blue in \Cref{fig:DOM}) contains a banner-related word.\footnote{We take extra precautions to filter out unlikely banner elements. For instance, if an element has a word from our corpus, but the element is set as \textit{invisible}, we discard the element (as the banner should be visible to the user). See Appendix \ref{app:Element_property} for details.} 

Next, it traverses up in the DOM hierarchy towards the HTML element that has either a \textit{positive z-index} or a \textit{fixed position} attribute. Generally, cookie banners are either displayed on top of the webpage content (positive z-index) or maintain the same position on the webpage (fixed position). The element with these properties very likely contains the banner. We call this the ``anchor'' element (see the green \texttt{<div>} element in \Cref{fig:DOM}). If \tool fails to find such an element, it considers the \texttt{<body>} element as the div anchor element.

The anchor element contains the banner, but the banner may still be fully contained within some sub-element of the anchor. To find this most-specific element, \tool traverses down again in the DOM, starting from the anchor element. It uses the following heuristic: the visible elements contained inside the anchor (\eg banner title, description, and buttons) should also be contained entirely within the more-specific candidate element.
Following this heuristic, \tool continues traversing down the DOM tree until it finds an element that does not completely contain all visible banner elements anymore. This implies that the parent of this element is the most-specific banner-containing element. This is shown as the red \texttt{<div>} element in the DOM tree in \Cref{fig:DOM}. %

Some websites might include banners as \texttt{iframes}, which are outside the regular website's DOM.
In cases where \tool fails to locate the desired element that contains the banner, it specifically iterates over all visible \texttt{iframes}. The above steps are once again repeated inside each \texttt{iframe} to detect the banner. 

\textit{Efficacy of banner detection:} We test our banner detection approach on the Tranco top-$1$k websites. We manually inspect and confirm that a total of $518$ websites show banners. Using \tool, we are able to correctly detect banners on $513$ websites. Therefore, only $5$ websites show a banner, but \tool fails to detect them. The reasons include the presence of a shadow DOM \cite{shadowdom} on the website (\texttt{godaddy.com}) and banners having words not present in our corpus (\texttt{washington.edu}). Similarly, only $4$ websites do not show any banner, but \tool incorrectly detects a banner.
For example, \texttt{allaboutcookies.org} has cookie-related words
in its DOM, but does not show a banner.
Overall, \tool detects banners with more than $99\%$ accuracy and extremely low FPR ($0.008$) and FNR ($0.009$).

\vspace{-1em}
\subsubsection*{\textbf{Banner Interaction:}}

After successfully detecting a banner, \tool can also interact with it. It can both ``accept'' and ``reject'' cookies in an automated manner. 
To do so, it relies on a corpus of words that are frequently used in cookie banners to indicate acceptance or rejection of cookies. This corpus consists of three categories of words indicating ``accept'', ``reject'', and ``settings'' (see Appendix \ref{app:corpus} for more details on how we create the corpus).
After successfully detecting the banner and identifying these words, \tool  automatically clicks the identified button. Throughout all our experiments we use three modes to interact with websites \ie ``No interaction'' (we do not click any button on the banner), ``Accept'' (we click accept related words), and ``Reject'' (we click reject related words). 

\tool first detects the banner and then identifies those HTML elements of the banner that contain words belonging to the said corpus. If it identifies multiple such elements, it first prioritizes \texttt{<button>} elements and then selects the one with the minimum number of words. For instance, in a banner, a \texttt{<p>} element may contain the text, ``To accept all cookies, please click the button below'', and another \texttt{<p>} element simply has the word ``accept''. Our tool selects the latter, as it is likely the button to provide consent.\footnote{One can simply detect the \texttt{<button>} tags and search for words inside them. However, we observe that banner buttons are not always implemented in this manner. Instead, many websites use other types of tags like \texttt{<input>} or \texttt{<div>} to implement buttons.}

To provide consent for cookies, \tool searches for words belonging to the ``accept'' category in the corpus. When it finds a match, it clicks on the identified element.
The process of rejecting cookies is similar: \tool searches for words inside the banner belonging to the ``reject'' category in the corpus.
However, if it fails to find such words, it attempts to reject the banner policies using the Never-Consent browser extension \cite{neverconsent}. Never-Consent searches for different functions provided by Consent Management Platforms (CMPs) to reject the banner policies (\eg OneTrust CMP's function \texttt{OneTrust.RejectAll()}).

However, if \tool still fails to reject the banner, it searches for the third category of words \ie ``settings''. This is because very often, the option to reject cookies is present inside a banner's settings. On a successful match, it clicks on the element containing the ``settings'' word. If the click is successful, the settings dialogue opens, and \tool again searches for words belonging to the ``reject'' category inside this dialogue. Using this approach, \tool can successfully detect, accept, and reject banners on websites. %

\textit{Efficacy of banner interaction:}
As previously mentioned, 518 websites of Tranco top-1K websites show a banner. We manually confirm that 444 of these offer an explicit accept option. %
The remaining 74 websites do not give the option to explicitly accept (\eg the banner just has a close button, or there is an implicit accept\footnote{Some websites show banners that do not overtly show the ``accept'' option. For instance banner on \texttt{bitly.com}, just states that ``By continuing to use this site you are giving us your consent to do this''.}). \df{\tool does not click accept button on any of these 74 websites.} 

In our research, we just consider explicit accept when interacting with banners. This is because, according to GDPR, websites must take users' consent explicitly. Later, for such websites, we quantify the increase in cookies after clicking the accept button.
With \tool, we successfully click accept on 430 out of 444 banners with explicit accept. However, amongst the remaining 14, \tool clicks the incorrect button on 13 websites. The banners of these websites contain buttons with words that negate the semantic meaning of accept, \eg ``NOT Accept'' (which is essentially a reject). 
Since \tool does not consider the text's semantics, it incorrectly classifies them as the accept. Lastly, only one website shows a banner with words that we do not have in our corpus. Thus \tool failed to click the button for that single website. Overall \tool successfully clicks the button with more than $97\%$ accuracy.

\df{Finally, we calculate \tool's reject accuracy by manually checking the screenshots for the Top-1k websites.
\tool successfully reject banners on 377 out of 524 websites and finds that 81 banners do not provide a reject option, resulting in an accuracy of 87.4\%.
The majority of unsuccessful rejections come from 38 websites that use multi-select mechanisms to reject cookies.}

\subsection{Cookie Classification}

Classifying cookies as first-party or third-party requires identifying the domain of the website as well as the received cookies. Thus, we use the public suffix list~\cite{public-suffix-list} to identify the domain of (1) the website and (2) the URL in the \textit{domain attribute} of the cookies. 
Then for each of the received cookies, we compare its domain with the website's domain. On a successful match, we classify the cookie as first-party; otherwise, we consider it a third-party. 

Next, similar to Götze \etal \cite{gotze2022measuring}, we use the justdomains blocklist \cite{justdomains} to identify tracking cookies.
This list contains entries from various popular tracking lists \viz EasyList, EasyPrivacy, AdGuard, and NoCoin filter lists, only if the \emph{complete domain} is identified as tracking. 
If the cookie domain matches one of the domains in the justdomains list, we classify it as a tracking cookie.
To ensure the correct classification of tracking cookies, we perform a small-scale validation:
We identify the top 100 websites sending the most tracking cookies and then we manually inspect the tracking cookie domain.
\df{We confirm that well-known tracking domains are indeed sending these cookies (\eg \texttt{doubleclick.net}).}

\subsection{OpenWPM Measurement Setup}
\label{app:measurements_platform}

We use Amazon EC2 instances in eight different geographic locations. 
These instances have four CPU cores and are provisioned with 16GB RAM.
For our measurements, we use \openwpm v0.18.0 \df{running Firefox} in stateless mode~\cite{stateless} with the following configuration.
In each run, we execute 7 browser instances in headless mode, with a 60s Selenium timeout\footnote{Selenium timeout indicates the duration that Selenium waits for a website to be loaded by the browser.}.
Empirically, we observe the vast majority of websites to be loaded within these 60s.
Moreover, we set the sleep time to 30s, which we experimentally find to be a suitable value. The sleep timer starts when the on-load event is triggered, ensuring that \openwpm remains on the website for this time period. This is necessary because some cookies are still being set even after the page has finished loading.
Furthermore, we set the \openwpm timeout\footnote{\openwpm timeout forces the current website crawl to stop upon expiration. That is useful, as Selenium freezes during the loading of some websites (\eg \texttt{bet365.com}).} to 360s (six times larger than the Selenium timeout).
\tool starts detecting the banner (and interacting with it if configured) in three attempts at 0, 10, and 20 seconds after the sleep time has started.
We see that more than 94\% of banners are detected just on the first try. To aid in manual verification of measurements, \tool takes a screenshot of the website before interaction, the detected banner, the clicked buttons, and the website after each click.

For the banner interaction measurements from the VP in Germany, which consists of 150,000 separate crawls (10k domains each with 5 repetitions and 3 different modes of interaction), 138,018 are reachable, 946 and 455 exceed Selenium's and \openwpm's timeout, respectively, for 10,175 the domain is unreachable, 406 trigger exceptions (\eg due to the lack of a \texttt{<body>} tag or page reloading during banner detection). In total, we consider 135,307 successfully completed measurements from all 8 vantage points in our analysis.

\subsection{Ethical Considerations}

Before conducting our Web measurements with \openwpm, we incorporate proposals by Partridge and Allman \cite{partridge2016ethical} and Kenneally and Dittrich \cite{kenneally2012menlo} and follow best measurement practices \cite{durumeric2013zmap}.
The AWS nodes are used only for measurement purposes, they are set up with informative rDNS names, they host a website with information about the measurements, and we offer the possibility for network administrators to be added to a blocklist.
We run \openwpm similarly as any regular user when visiting websites with a normal browser.
During our measurement period, we did not receive any complaints.

\section{Effect of Cookie Banners}
\label{sec:effectsbanners}

As  most research involving GDPR does not consider banner interactions (\ie clicking accept/reject buttons) \cite{acar2014web, englehardt2015cookies, trevisan2019,linden2020privacy}, we develop \tool to automatically interact with banners.

\begin{figure}[t!]
\centering
\includegraphics[width=0.9\columnwidth]{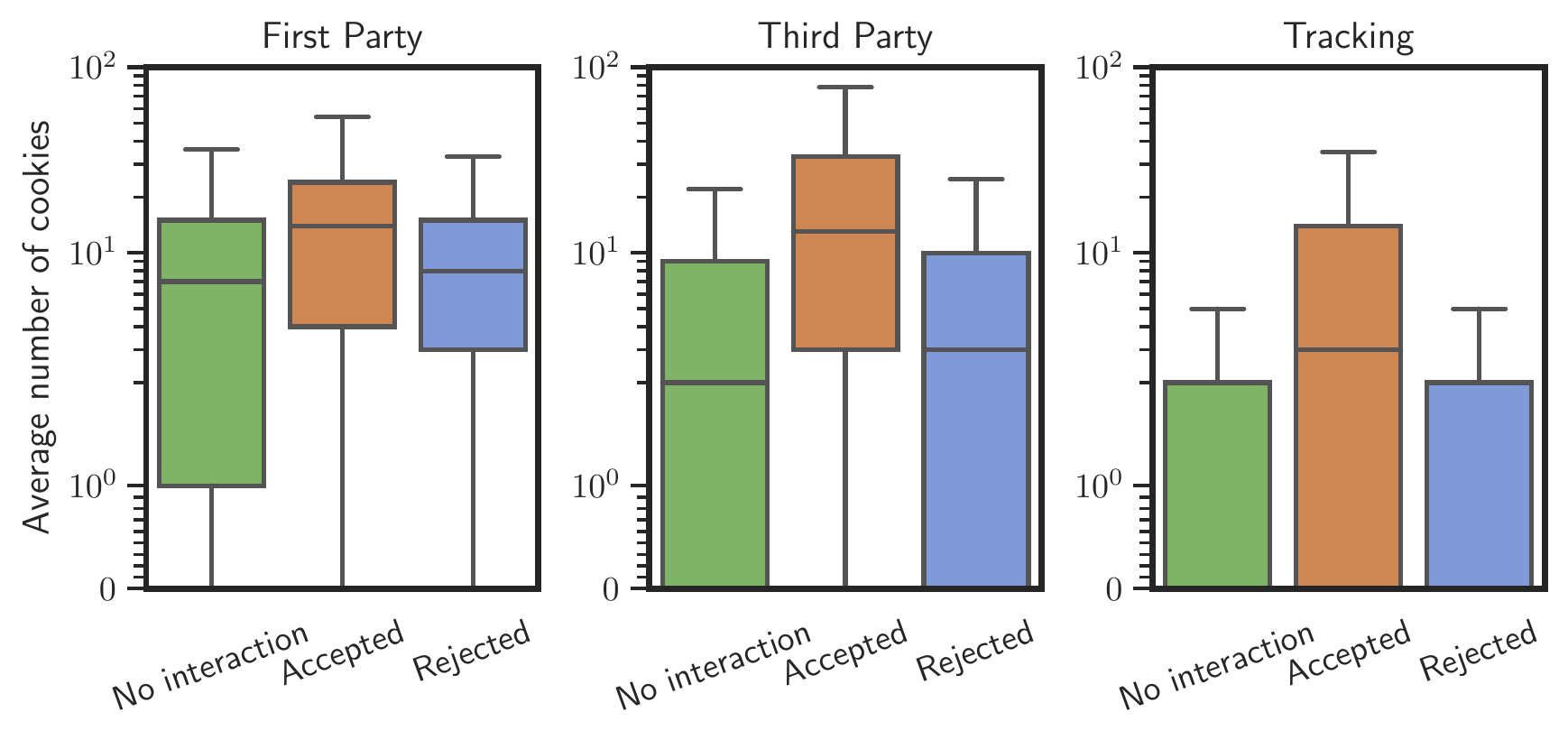}
\caption{Cookie differences between no interaction, accept, and reject from the Germany VP.}
\label{fig:banners}
\end{figure}

\begin{figure*}[ht!]
\centering
\includegraphics[width=0.9\textwidth]{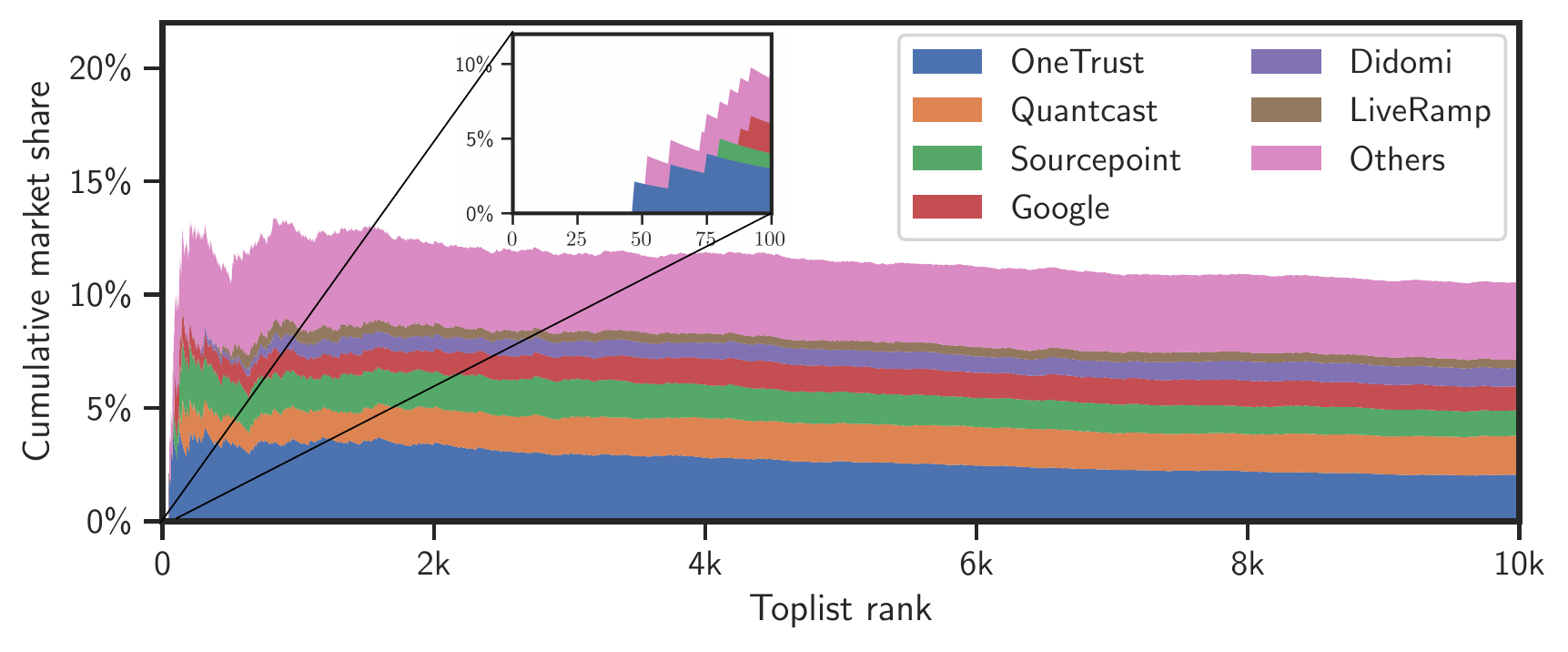}
\caption{CMP distribution depending on the Tranco rank from the Germany VP.}
\label{fig:cmps}
\end{figure*}

We run \tool on the Tranco top-10k websites \cite{Tranco} to analyze the effect of cookie banners.
First, we investigate how many websites we can detect and interact with banners.
From the vantage point in Germany, we can successfully detect banners on about 47\% out of all accessible websites.
\tool is then able to click on Accept and Reject buttons of the banner for around 40\% and 30\% of all websites, respectively.
Next, \Cref{fig:banners} shows how interacting with banners can substantially impact cookie distribution. After accepting a banner, the number of first-party (FP) cookies increases by more than $1\times$ and the number of third-party (TP) cookies increases by $5.5\times$ on average. As for tracking cookies, the average increase from zero to 7 which shows a significant impact. %
Also, the minimum number of cookies set by 75\% websites (lower quartile) increases from 1 to 4 and 1 to 3 for FP and TP cookies, respectively; for tracking cookies it remains 0. 
Moreover, we observe a jump in the maximum number of cookies set, which for third-party cookies, and consequently tracking cookies, is quite noticeable.
As for the rejection impact on first-party cookies, we can also see a slight increase in the number cookies. This might be because of cookies that are being set to keep the state of rejection for future website access.
This is further corroborated, as we do not see this trend for third-party cookies.
Furthermore, we see that the number of tracking cookies is quite low (near zero) when the banner is not accepted, which indicates the effectiveness of GDPR to reduce tracking.
Overall, we find that banner interaction has a large influence on the number of cookies, and it is therefore imperative to use tools like \tool to take banner interactions into account.

While accessing these websites with \tool, we also analyze the distribution of 
Consent Management Providers (CMPs).
CMPs are platforms that offer cookie consent handling as a service, \ie websites can include a ready-to-use, yet configurable banner instead of developing their own cookie banner solution.
The IAB Europe Transparency and Consent Framework (TCF) is a GDPR-compliant consent solution that specifies the overall behaviors of CMPs \cite{tcf}.
As mentioned in the specification of TCFv2 \cite{tcf2}, all CMPs need to implement a \texttt{\_\_tcfapi()} function which allows third parties to have access to the users' selected preferences and act accordingly.
In \tool we use this function to record the name of the CMP while crawling a website.
We observe that---contrary to the specification---not all websites with CMP banners actually implement the \texttt{\_\_tcfapi()} function.
This specification violation is not limited to a specific CMP.
To obtain a better and more comprehensive distribution for CMPs, we additionally incorporate results from the Never-Consent browser add-on \cite{neverconsent} into our data.
Never-Consent leverages custom APIs which some CMPs implement in addition or instead of \texttt{\_\_tcfapi()}.
These custom APIs allow for interaction with CMPs to fetch user-related data or can even trigger a \emph{reject all} event.

In \Cref{fig:cmps} we show the cumulative market share of different CMPs for the Tranco top-10k websites.
As we can see, in total within the top-1k websites around 13\% of websites use CMPs.
The CMP deployment remains almost constant with increasing rank, hinting at a consistent CMP deployment between ranks 2k and 10k.
The CMP ecosystem is dominated by four companies (OneTrust, Quantcast, Sourcepoint, and Google) which are responsible for more than half of all CMP banners.
Interestingly, we can not find a single website in the top 46 websites using a CMP and there is a generally much lower CMP deployment among top-ranked websites (see zoomed-in figure).
This can be attributed to the fact that large Internet companies tend to avoid relying on third parties for handling privacy-sensitive data.

Throughout our study, we see a slight increase in CMP usage: From 95 websites out of the top-1k in %
July 2021 to 107 websites in January 2022.
Therefore, it seems that CMPs will continue to play an important role in the cookie ecosystem, which future research should take into account.

As for other VPs, we see fewer CMPs detected on average. This is due to some CMPs not implementing their APIs (\ie \texttt{\_\_tcfapi()} or custom ones), when they do not show a banner, which happens more for non-EU VPs. There is also an increase in the share of CMPs in the category ``Others'', which underlines that popular CMPs are less likely to provide APIs if no banner is shown.

Finally, we also compare our CMP results to previous work \cite{hils2020measuring}.
Their results for CMPs following TCFv1 are similar to our results for the new TCFv2 standard.

\section{Impact of Geographical Location}
\label{sec:location}

\begin{wrapfigure}{l}{.5\textwidth}
\centering
\includegraphics[width=.5\textwidth]{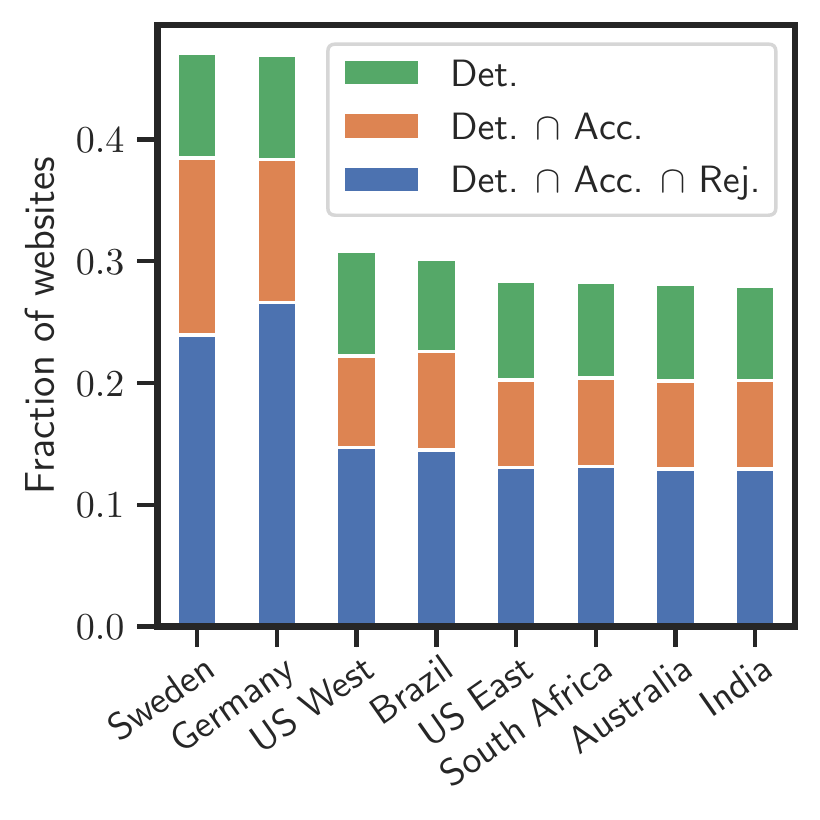}
\caption{Effect of location on banner detection, accept, reject.}
\label{fig:banner-location}
\end{wrapfigure}

We examine the effect of geographical location on banner interaction and Web cookies to observe if websites behave differently (\eg set a different number of cookies) in different regions. We crawl the Tranco Top-10k websites from eight geographically diverse vantage points (see \Cref{subsec:locations} for more details). While accessing the websites, we interact with the banners in three modes: no interaction, accept, and reject. \Cref{fig:banner-location} depicts the impact of geographic location on banner detection and interaction. In non-EU countries, we detect banners on less than 30\% of websites, whereas, in EU countries we observe more banners (\ie on about 47\% of the Tranco top-10k).
This is a banner increase of 56\% from non-EU compared to EU.
Also, across all locations, \tool is able to accept more banners (blue + orange) than reject them (only blue).\footnote{The slightly lower number of rejects in Sweden compared to Germany is due to a lack of Swedish reject-related words in our corpus.}
This indicates that banners are biased towards showing more accept than reject options.

To analyze the effect of geographical location on cookies, we visit each website five times in each mode and record the number of cookies. \df{If a website is not accessible in five of the iterations at any location, we exclude it from our geographical location analysis.}
We now report the cookie trends observed at different locations in different modes.

\noindent \textbf{No Interaction Mode:} 
In the no interaction mode, 63\% of sites set a different number of TP cookies in at least one location. Of these websites, 56\% follow a trend where they set the highest number of TP cookies in either the US East or the US West and the least in Germany and Sweden.
We also confirm that in the EU region, about 56\% of websites set TP cookies and 30\% set tracking
{%
\parfillskip=0pt
\parskip=0pt
\par}
\noindent cookies even in the no-interaction mode. In non-EU regions, a larger proportion of websites set TP (64\%) and tracking cookies (43\%).
\df{This indicates that GDPR has a positive impact on the reduction of TP and tracking cookies, but still many websites set these cookies without the users' consent. Setting TP (especially tracking) cookies before taking users' approval is a clear violation of GDPR.}

\noindent \textbf{Accept Mode:}
When analyzing the accept mode, we focus on those websites where we can successfully detect and accept banners at all VPs (\ie 18\% of Tranco top-10k). This ensures that banner presence and different banner languages due to varying VPs do not influence our analysis.
Amongst them, 21\% of websites send precisely the same number of TP cookies at all locations; examples include \texttt{truecaller.com}, \texttt{ghostery.com}, and \texttt{deepmind.com}. These websites represent an ideal case where users from different regions receive the same number of TP cookies after consenting to the banner. This is noteworthy as even users who reside in regions without strong data protection laws (\eg India) experience similar privacy standards to those that live in the regions protected by such laws (\eg EU).

To further assess the impact of GDPR on TP and tracking cookies, we now consider those websites that offer banners \textit{only} in the EU and on which \tool is able to click the accept button (\ie 37.6\% of the total). 
For such websites, we observe that the variation in TP cookies is nearly identical for both VPs in the EU. We find a similar trend across the rest of the VPs in non-EU regions. Thus, we aggregate the data points per website for VPs in the EU, and separately for all non-EU ones.

\begin{figure}[h!]
\centering
\includegraphics[width=0.9\textwidth]{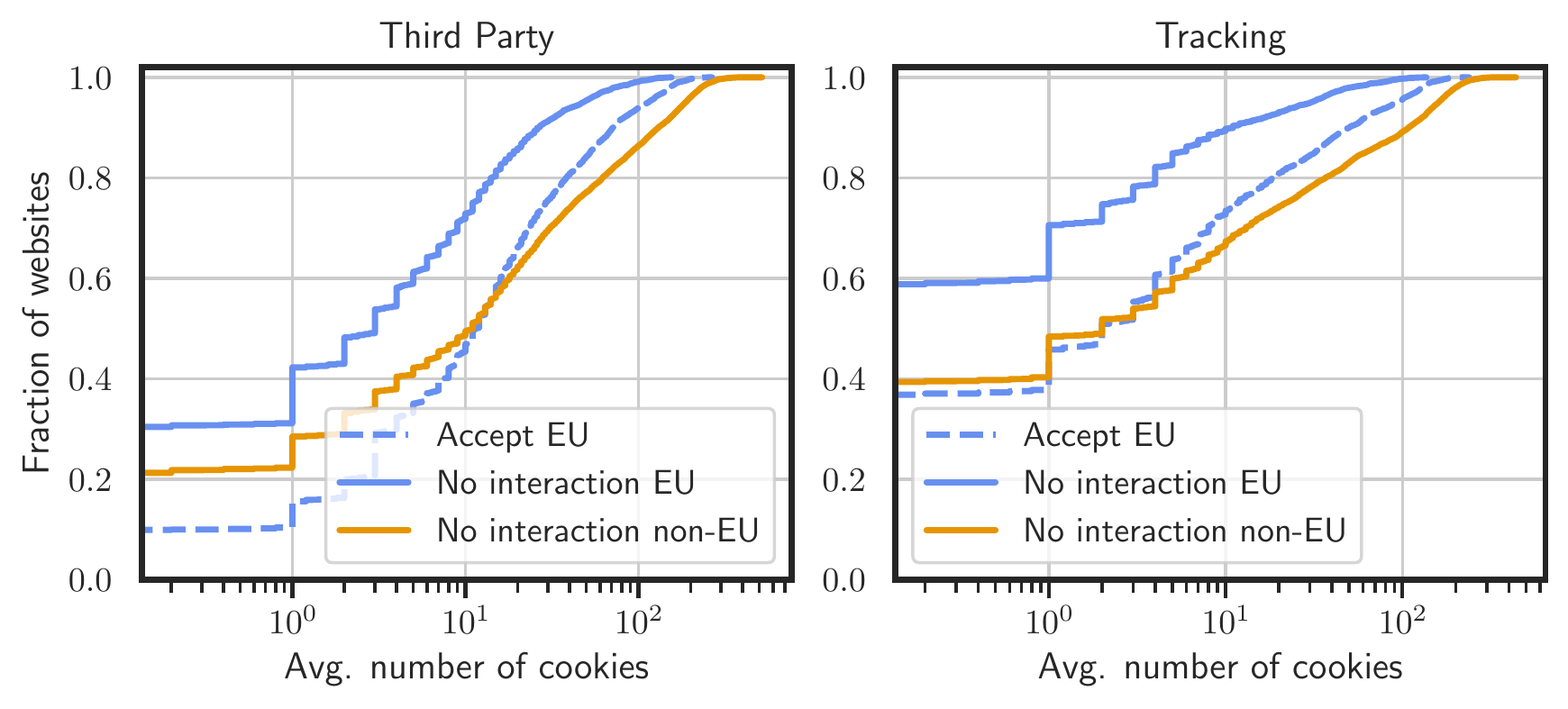}
\caption{ECDF plot with the average number of TP (left) and tracking (right) cookies for websites on which \tool is able to click accept only in the EU.}
\label{fig:tp-cookies}
\end{figure}

In \Cref{fig:tp-cookies} we show an ECDF of the number of TP and tracking cookies for both EU (in blue) and non-EU regions (in orange). It is evident from the figure that, before interaction, about 60\% of websites in the EU region set, on average at most 5 TP cookies, and about 80\% of websites set, on average at most 4 tracking cookies.
On the contrary, in non-EU regions, 60\% of the websites set at most 20 TP cookies, and 80\% set at most 40 tracking cookies \ie an increase compared to the EU region by a whole magnitude.
Interestingly, 65\% and 83\% websites set fewer TP and tracking cookies respectively, even after accepting the banner policies in the EU, compared to no interaction at non-EU VPs. This shows that GDPR has a noticeable impact on the number of TP cookies. However, as expected, we find that GDPR does not impact FP cookies: 70\% of websites set more or an equal number of cookies after accepting the banner compared to no interaction at the non-EU VPs.

\noindent \textbf{Reject Mode:} For the reject mode analysis, we again select websites that again show banners only in the EU, and for which we are able to click the reject button (\ie 23.7\% of the total).
We find that 87\% and 96\% of these, set fewer TP and tracking cookies respectively in the EU after rejecting the banner compared to the no interaction mode at non-EU VPs.
We observe a similar trend for FP cookies: 72\% of these websites set fewer FP cookies in the same scenario.

Overall, our results indicate that GDPR has a positive impact on reducing the number of TP and tracking cookies, but we do not find any measurable effect of other privacy laws (\ie LGPD and CCPA) on TP and tracking cookies. This observation holds good for banner detection as well; we detect a maximum number of banners in the EU countries.

\section{Website Cookie Consistency}
\label{sec:consistency}

Next, we analyze the consistency of website cookie behavior, in order to learn how consistently websites send a certain number of cookies.
This is important to ensure, that what we measure is not influenced by website randomness, \ie due to excessively changing third-party content.
For statistical consistency analysis, we visit each website of the tiered Tranco top-10k (100 websites each in three different rank tiers) 100 times for each of the three different interactions (no banner interaction, accept, reject).

\p{Intra-location consistency:}
To draw meaningful conclusions about cookie characteristics, one must ensure that a website sends a similar number of cookies when accessed multiple times from the same location. \Eg if a website, when accessed for the first time, sends only five cookies, but when accessed the second time, sends hundreds of cookies, it should be classified as inconsistent. For such websites, it is non-trivial to draw meaningful conclusions from the measurements.

\begin{figure}[t!]
	\centering
	\includegraphics[width=0.9\textwidth]{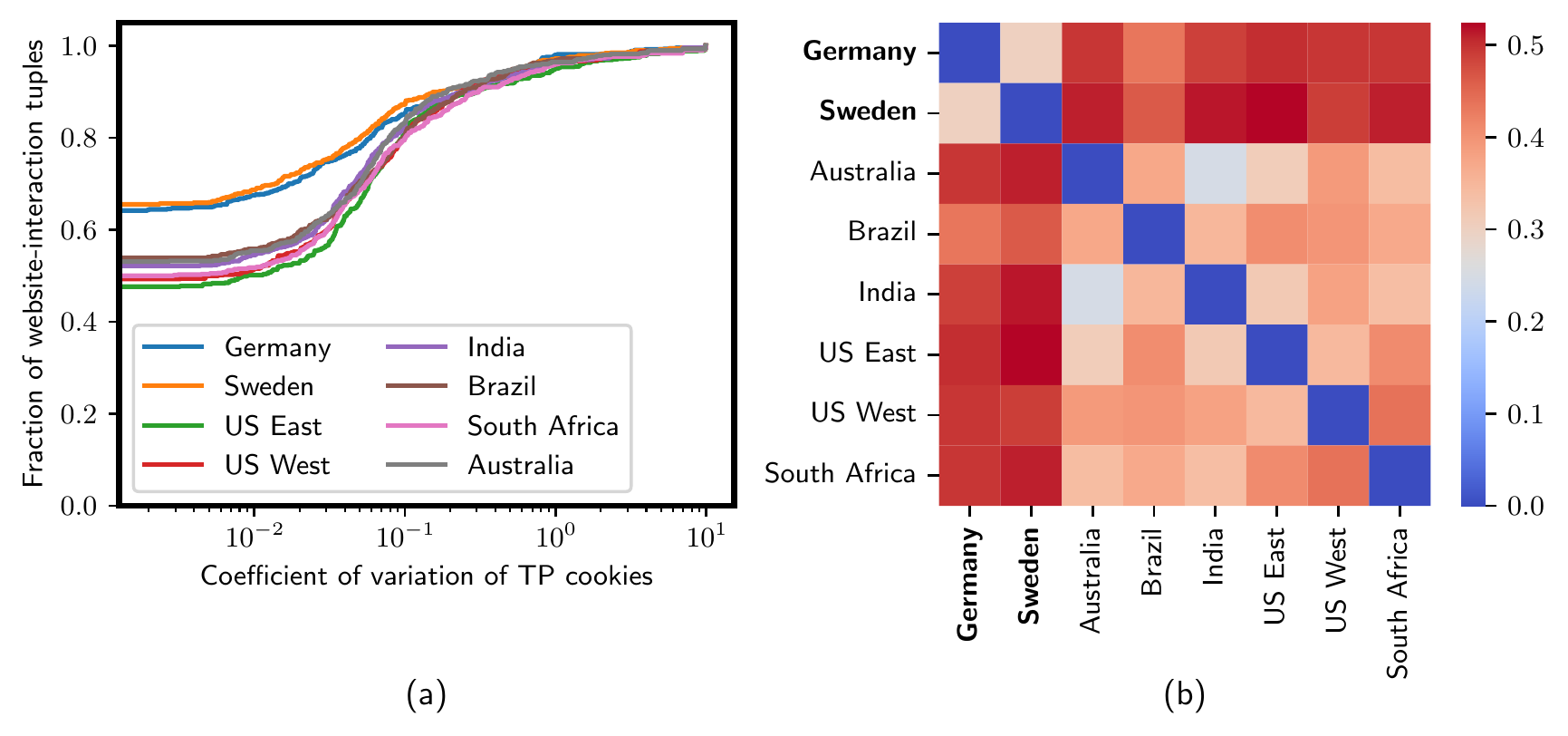}
	\caption{(a) Intra-location consistency of third-party cookies. (b)  Inter-location statistically significant differences of third-party cookies (EU VPs in bold).
}
	\label{fig:CoV_TP}
    \end{figure}

From each of the VPs (in eight countries), we measure the intra-location consistency using the coefficient of variation (CoV) as a metric.
The CoV is calculated by dividing the standard deviation by the mean.
The smaller the CoV, the more consistent the cookie behavior is, when looking at it from each VP separately.
We visit each website of the tiered Tranco list from each location and then calculate the CoV based on the number of cookies the website sends.
\Cref{fig:CoV_TP} (a) shows the ECDF of CoV for third-party cookies.
We can clearly see two groups of websites in the plot: EU (Germany and Sweden) on the top and non-EU below that.
It seems that when visiting websites from within the EU, they exhibit a more consistent cookie behavior.
However, this difference is influenced mainly by the number of websites that send exactly zero third-party cookies which result in a CoV of zero:
More websites when visited from within the EU send exactly zero third-party cookies, compared to when visited from a non-EU VP.
This in turn leads to the ECDF curves of EU countries starting higher than non-EU countries, exhibiting a shifted, but the similar curve and later even merging.
This is another indicator of the effect of the VP's geographical location in combination with GDPR on cookie behavior, as pointed out in \Cref{sec:location}.
Overall, we find that 75--80\% of websites are consistent with a CoV of less than 0.1 (\ie the standard deviation is at most 10\% of the mean).
For first-party cookies (not shown) we see a more similar picture across VPs.

\p{Inter-location consistency:}
To find statistically significant differences in the number of observed cookies depending on the VP location we use the Mann-Whitney U (MWU) test~\cite{mann1947test}\footnote{\df{The MWU test is a statistical post hoc test, \ie it allows to find differences in the cookie distribution between all pairs of VP locations.} Our setup fulfills the MWU assumptions, \ie all test samples from both groups are independent of each other, the samples are ordinal. The distributions of both populations are identical under $H_0$ and not identical under $H_1$.}.
Again, we crawl websites from the tiered Tranco list 100 times for each interaction (no interaction, accept, reject) from each VP. 
Then we apply the MWU test with Holm p-value correction \cite{holm1979simple} and choose a p-value of 0.05 to determine statistical significance.
In \Cref{fig:CoV_TP}(b) we show a heatmap depicting the statistical differences. In the figure, we see two main clusters, \ie EU vs. non-EU  and non-EU vs. non-EU. We find that the majority of differences occur between EU (bold label) and non-EU locations, with more than half of all website-interaction tuples showing a statistically significant difference.
On the other hand, if both locations are either in the EU or both outside the EU, we see fewer differences. Moreover, we also confirm that the Tranco rank tier does not affect the differences.
\df{An example of such a website is \texttt{nytimes.com}, which sends on average 5 TP cookies when visited from Germany or Sweden, 10 TP cookies from Brazil, and more than 80 TP cookies from other countries.}

In conclusion, when visiting a website from a GDPR country compared to a non-GDPR country, there is a significant difference in third-party cookies being sent by most websites.
For first-party cookies (not shown) we see a similar picture across VPs, although with fewer differences in total.

\section{Landing vs. Inner pages}
\label{sec:landVsInner}

When users access a website, they often not only access the website's main landing page but navigate through other inner pages of the website as well. For instance, people visiting the landing page \texttt{https://www.bbc.com/} could access the article on the inner page \texttt{https://www.bbc.com/sport/football/58920223}.
Thus, it is important to study the differences between cookies for landing and inner pages for a given website.
We use a simple criterion to classify a link as an inner page (corresponding to a given landing page). An inner page link must begin with the landing page's fully qualified domain name (FQDN). For instance, \textbf{https://www.bbc.com/}sport/football/58920223 is the inner page of \textbf{https://www.bbc.com/}.

We intended to use the Hispar list \cite{aqeel2020landing} (which contains links to seemingly inner pages) for our analysis. However, we find that many inner pages mentioned in the list either do not begin with the FQDN of the landing page or redirect to completely different domains. For instance, \texttt{mail.google.com} is classified as an inner page of \texttt{google.com}, which in practice it is not. In general, we observe that more than $50\%$ of inner pages (corresponding to a landing page) in the Hispar list are actually not inner pages. %
Thus, we use our own automated approach to access a given website's landing and inner pages. For our analysis, we select $10$ random inner pages for each landing page as follows. 

We first access the landing page of the given website (\eg \texttt{https://www.bbc.\\com/}). The obtained HTML page contains Web links to inner as well as non-inner pages. Next, we select a link by crawling for \texttt{<a>} elements and check whether it is a potential inner page or not. As already mentioned, we simply check that the inner page link must begin with the landing page’s FQDN. Using Selenium, we visit this link and extract the final link (which might have changed due to redirection). If the link is an inner page, we append it to the list of inner pages. If the link is already present in the list, we ignore it and proceed with the remaining ones. Finally, we stop searching for inner pages when either $10$ inner pages are found or a total of $50$ links (present on the landing page) have been tested. We repeat the same process for all tiered Tranco websites.

\begin{figure}[t!]
\centering
\includegraphics[width=0.9\textwidth]{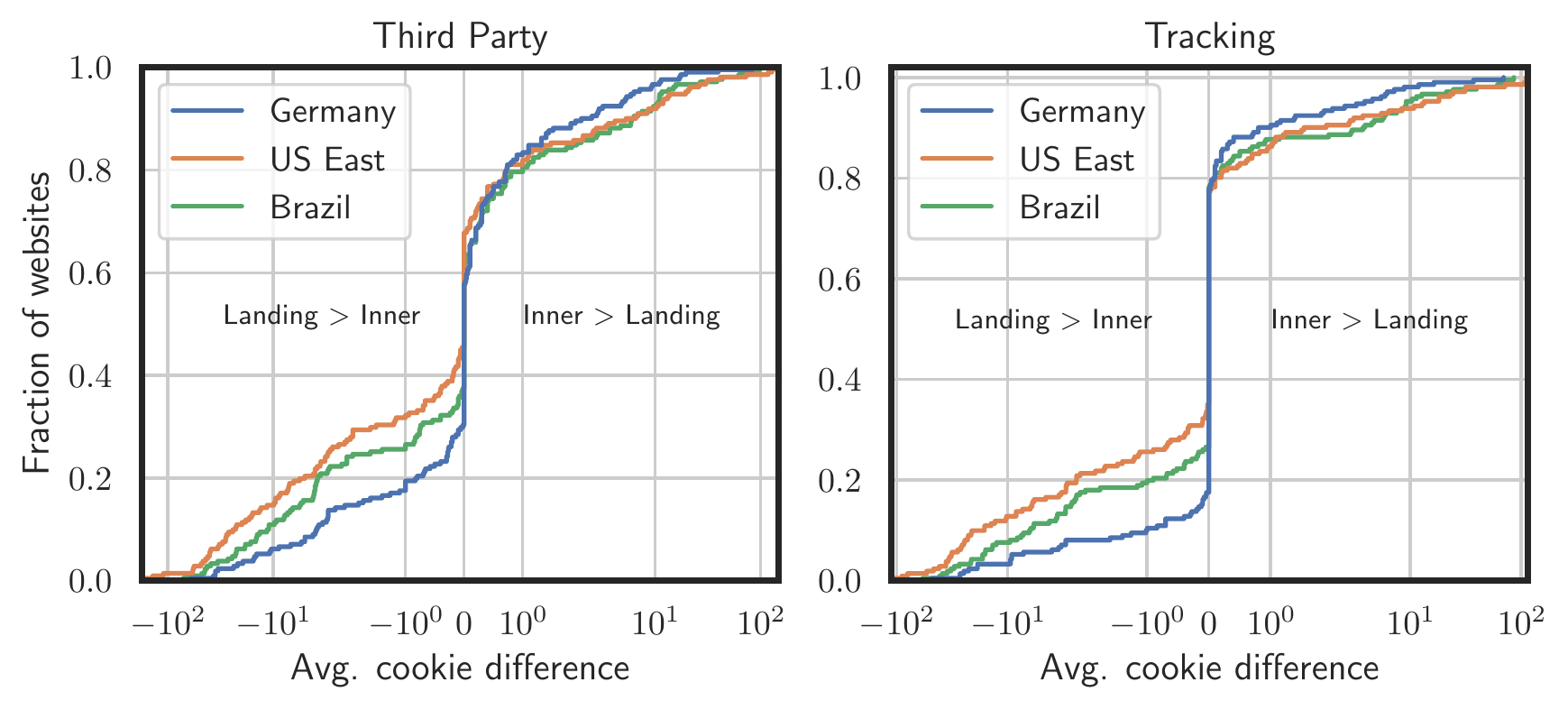}
\caption{Average number of TP cookies comparing landing vs. inner pages.}
\label{fig:tp-cookies-inner}
\end{figure}

In total, we obtain 2273 inner pages corresponding to 300 Tranco websites.
We access the set of landing and inner pages from all VPs. Like our other experiments, we visit each webpage (landing and inner) five times in each mode (no interaction, accept, reject) and record the average number of cookies per webpage. \Cref{fig:tp-cookies-inner} shows the ECDF of the difference of average TP and tracking cookies from the ten inner pages compared to the corresponding landing page (in the no interaction mode). The negative difference on the x-axis (left part of the figure) corresponds to the fraction of websites where we observe more cookies on a landing page than on inner pages (shown as Landing $>$ Inner). Zero means the same number of cookies is found for both categories. Positive values (right part of the figure) correspond to the fraction of websites where more cookies are sent on inner pages than the landing page (represented as Inner $>$ Landing).
\Cref{fig:tp-cookies-inner} depicts this difference for three VPs \ie US East, Brazil, and Germany. We show only these three VPs because we observe nearly the same trend for US East and US West; observations in Brazil are quite similar to India, South Africa, and Australia; the trend in EU countries is almost the same.

At all of our VPs, we find that 12.7\% and 8\% of websites set more TP and tracking cookies, respectively, on the landing page than on the inner page (\eg \texttt{amazon.com}, \texttt{vk.com}, and \texttt{youtube.com}). Looking at VPs separately, the proportion of such websites is the highest in US East (32\% TP and 24\% tracking) and the lowest in Sweden (21\% TP) and Germany (12.3\% tracking).
Moreover, our analysis reveals that 87\% of these websites set at least 10 more TP cookies on average on the landing page at all locations. One possible explanation for this trend could be that many websites show more content on the landing page, include more third-party content, and thus set more TP cookies.

Similarly, we observe that 14.7\% and 7.7\% of websites set more TP and tracking cookies respectively on inner pages across all VPs (\eg \texttt{cnn.com}, \texttt{bbc.com} and \texttt{reddit.com}).
When investigating each VP separately, the proportion of such websites is the highest in Germany (29.7\% TP) and South Africa (19.3 tracking), and the lowest in US East (22\% TP) and Brazil (15.3\% tracking). It is interesting to note that, although GDPR discourages the use of third parties without consent, a substantial fraction of websites prioritize setting TP cookies on inner pages. 
This could also facilitate user profiling \cite{acar2014web} as third-party services could better characterize users' viewing habits and choice of content at a more fine-grained granularity.
Overall, our results indicate that studying \textit{only} the landing page provides a partial picture of the TP cookies a user might get. In total, 49.3\% and 27.3\% of websites set a different number of TP and tracking cookies respectively on landing and inner pages at all our VPs.

\textit{Banners on inner pages:}
We check for banner presence as a potential contributing factor.
\df{Although we find a small number of websites with different banner behavior (\eg \texttt{www.colorado.edu/map}), we generally see a similar number of banners on landing and inner pages.}
Overall, using \tool, we detect banners on 22\% (US East), 51\% (Germany), and 30\% (Brazil) of the landing pages of the tiered Tranco list. Correspondingly, we detect banners on 25\% (US East), 50\% (Germany), and 31\% (Brazil) of the  inner pages.

\section{Mobile vs. Desktop}
\label{sec:mobilevsdesktop}

We look into the effect of visiting websites from browsers in desktop vs. mobile environments to understand how websites and third parties behave in this context. To visit a website from a mobile browser, we modify the default \openwpm user agent\footnote{Desktop: ``Mozilla/5.0 (X11; Linux x86\_64; rv:95.0) Gecko/20100101 Firefox/95.0''; mobile: ``Mozilla/5.0 (Android 12; Mobile; rv:68.0) Gecko/68.0 Firefox/93.0''.} and the screen size\footnote{Desktop: 1366x768; mobile: 340x695.}. We manually confirm that modifying these parameters change the appearance of most websites\footnote{In some cases this also changes the URL, \eg by prepending \texttt{m.} or \texttt{mobile.} to the domain name.} and we see both desktop and mobile versions of the same website. Interestingly, even with these minimal changes, we observe substantial differences between measurements conducted from desktop vs. mobile. We crawl the 300 tiered Tranco websites 5 times in each mode of interaction from all VPs with desktop and mobile configurations. 

\begin{figure}[h!]
\centering
\includegraphics[width=0.9\textwidth]{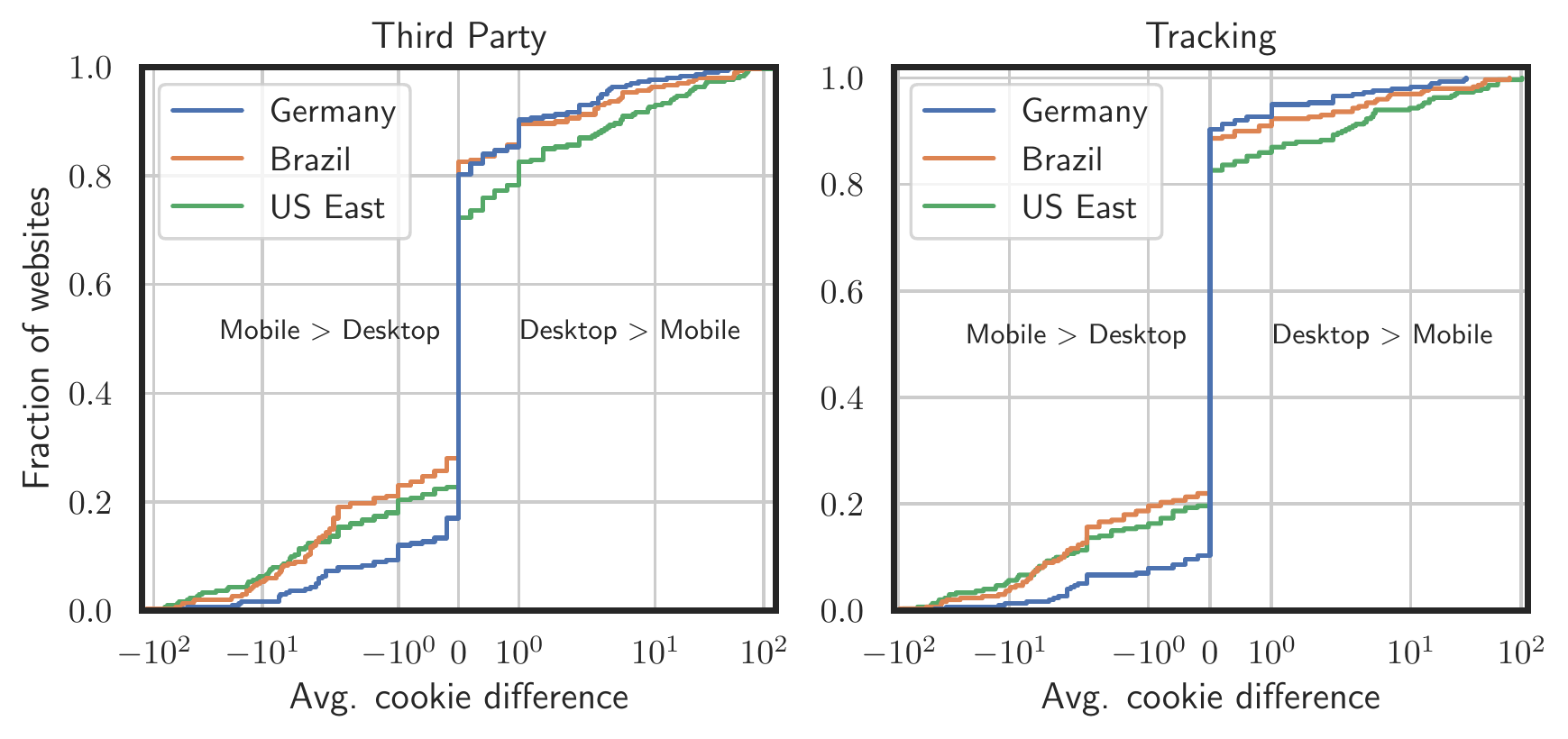}
\caption{Average number of TP cookies comparing mobile vs. desktop.}
\label{fig:tp-cookies-mobile}
\end{figure}

\Cref{fig:tp-cookies-mobile} shows the difference between the average number of TP and tracking cookies measured per website when visited from a browser on desktop vs. mobile in the no interaction mode.
We subtract the number of cookies observed on mobile from what we observe on desktop.
Hence, websites that set more cookies on the desktop yield a positive cookie difference on the x-axis.
Vice-versa, if a website sets more cookies on mobile, the cookie difference is negative on the x-axis. We observe that the TP and tracking cookies variation is nearly the same for US East and US West. The data from the VPs in the EU are alike, and the data from the remaining VPs are similar to each other. Hence, we plot the TP and tracking cookies per website for US East, Germany, and Brazil representing their respective classes.

At all VPs, we find that 7.3\% and 2.7\% of websites set more TP and tracking cookies, respectively, when visited from a desktop (\eg \texttt{bing.com}, \texttt{twitch.tv}).
On investigating VPs independently, we find that the proportion of such websites is the highest in US East (28\% set TP and 17\% set tracking cookies) and the lowest in Brazil (17\% set TP cookies) and Sweden (9\% set tracking cookies).
From our analysis, we note that 7\% of websites set at least 10 more TP cookies when being visited from a desktop from US East. These facts can be attributed to some websites having more content and hence more embedded third parties on desktop than on mobile. Many websites, when designed for mobile, decrease the number of advertisements and limit the content to what is visible without scrolling. This reduces data usage and improves the user's viewing experience.

We also observe that 7.3\% and 6.3\% of websites set more TP and tracking cookies, respectively when viewed from the mobile environment across all VPs (\eg \texttt{nytimes.com}, \texttt{livestream.com}). Distinct VP analysis shows that the proportion of such websites is the highest in Brazil (28\% set TP cookies, 22\% set tracking cookies) and the lowest in Sweden (15\% set TP cookies) and in Germany (10\% set tracking cookies). Our analysis shows that 4\% of websites set at least 10 more cookies when visited from mobile from non-EU VPs. As users are increasingly spending more time on their mobile devices \cite{mobile-desktop}, some third parties seem to be prioritizing placing more cookies when sites are visited from mobile for better targeting. It becomes imperative that measurements from mobile environments also be considered for a real-world analysis of cookies.

Overall, we observe that 14.6\% 
websites set a different number of TP cookies when accessed from desktop and mobile environments at all our VPs. Furthermore, our findings show a higher degree of similarity between desktop and mobile compared to previous work \cite{yang2020comparative}, which did not consider banner detection or interaction at all.

\textit{Banners on websites browsed from mobile:}
We check for banner presence as a potential contributing factor in this experiment as well. Using \tool we detect a similar number of banners on  websites when visited from desktop and mobile ($\approx$ 21\% US East, 46\% Germany, and 26\% Brazil).

\section{Impact of CCPA}
\label{sec:CCPA}

The California Consumer Privacy Act (CCPA) came into effect in January 2020.
In the context of CCPA, selling personal information in the form of TP cookies has been a widely debated topic \cite{TP-cookies-debate}.
Thus, we take the first step to studying how CCPA-compliant websites deal with third-party cookies.
To analyze the cookie landscape of such websites, we first need to find which websites are overtly complying with CCPA. For this, we use a straightforward approach. Websites covered by CCPA must include a conspicuous hyperlink on their homepage with the text ``Do Not Sell My Personal Information'' (DNSMPI) \cite{van2022setting}. We crawl the tiered Tranco list and identify websites that contain this hyperlink.\footnote{We use $8$ different phrases for searching DNSMPI hyperlinks (\eg ``do not sell my info'') as suggested by Van Nortwick \etal \cite{van2022setting}.}

Out of 300 tiered Tranco websites, we identify that 39 websites contain DNSMPI links from our US West vantage point, 29 websites from US East, and 21 from Germany. 
This indicates that a user's location impacts whether or not the DNSMPI link is shown. Interestingly, this applies to different locations within the US as well, \ie we see 11 websites that only show the DNSMPI link to clients from California but not when visiting the website from the US East.

To observe the impact of CCPA on TP cookies, we compare the TP cookies of websites containing DNSMPI links with websites that do not include said links. We select our US West (\ie California) VP for this analysis.
First, we classify the 39 websites with DNSMPI links into three sets belonging to Tranco top-100, 1001--1100, and 9901--10k, respectively. For instance, we obtain 12 websites that belong to the first set. Thus, to have a fair comparison, we randomly select the same number of websites without a DNSMPI link from the Tranco top-100 websites only. We repeat the same process for the other two sets as well. In the end, we compare websites in the same Tranco rank tier. In total, we compare 39 websites with DNSMPI links with the same number of websites without DNSMPI links.
This approach ensures that differences in TP cookies are not due to differences in Tranco rank.

\begin{wrapfigure}{L}{.5\textwidth}
\centering
\includegraphics[width=.5\textwidth]{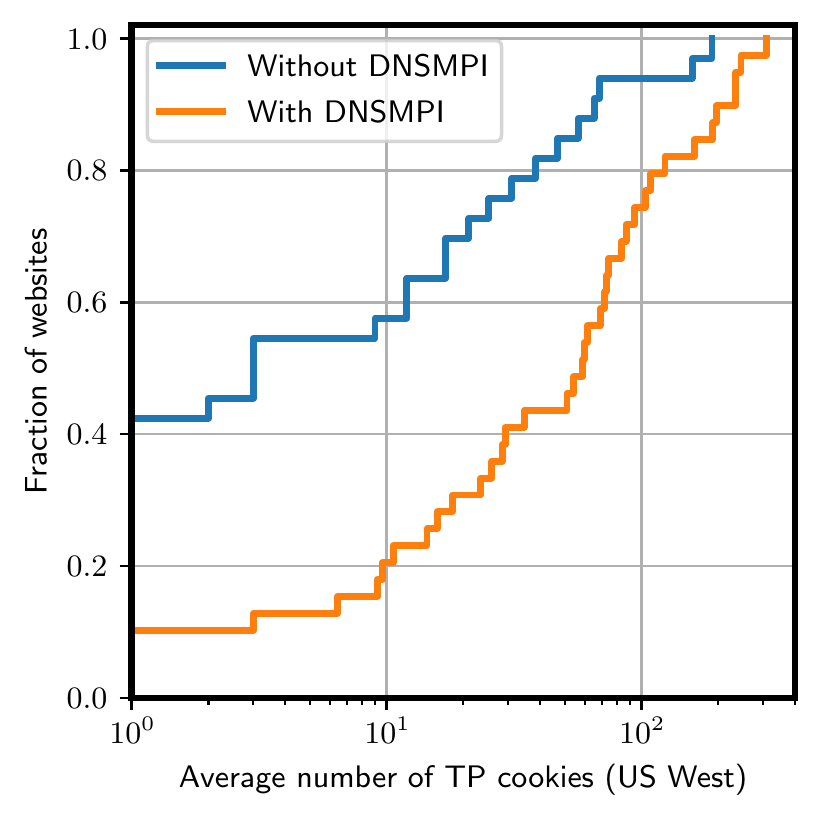}
\caption{Effect of CCPA on cookies: Websites with DNSMPI links send more TP cookies.}
\label{fig:tp-cookies-ccpa}
\end{wrapfigure}

Similar to previous experiments, we crawl each website five times and record the number of TP cookies.
\Cref{fig:tp-cookies-ccpa} illustrates the variation in average TP cookies for DNSMPI and non-DNSMPI websites (without cookie banner interaction).
We can see that websites without DNSMPI (blue line) set a lower number of
{%
\parfillskip=0pt
\parskip=0pt
\par}
\noindent TP cookies than the websites with DNSMPI (orange line). For example, $42\%$
of non-DNSMPI websites set on average just two or fewer TP cookies, whereas the same fraction of DNSMPI websites send 30 or fewer cookies. For tracking cookies, the trend is the same as TP cookies.

We further extend our analysis to Tranco top-10k websites, where we identify a total of 1373 websites with DNSMPI links from the US West. We observe a similar trend as we see with the tiered Tranco list. This shows that CCPA does not have a positive impact on TP cookies by default.
On the contrary, websites overtly adhering to CCPA send, on average more TP cookies than non-DNSMPI websites.
Furthermore, users need to manually look for the often well-hidden (\eg in website footers) DNSMPI links and click them to get any real benefit.
When it comes to reducing the number of cookies, CCPA seems much less effective than GDPR or similar legislation.

We check if banner presence could be contributing to the TP cookie differences for DNSMPI and non-DNSMPI websites.
To our surprise, we find that DNSMPI websites are twice as likely to show a banner compared to non-DNSMPI websites.
As a result, DNSMPI websites show a banner more often but still send more TP cookies.
\section{Discussion}
\label{sec:Discussion}

\noindent\textbf{Cookie banner automation:}
Since GDPR~\cite{GDPR} and similar privacy legislation came into effect, cookie banners have become more and more prevalent on the Web.
Moreover, during our measurements, we also see a wide variety of different banners.
This not only makes automated detection and interaction more challenging for research purposes, but it also hinders browser and extension developers to effectively interact with banners in an automated fashion.
These often rely on manually curated rules, do not have the option to reject cookie consent \cite{idontcareaboutcookies}, or are no longer maintained \cite{neverconsent}.
Efforts to offer a general easy-to-use mechanism to refuse all tracking cookies such as HTTP's ``Do Not Track'' header \cite{dnt}, have not been adopted by the advertising industry and were therefore abandoned.
The deployment of Consent Management Platforms (CMPs) could be leveraged as a standardized API for application developers to automate banner interaction.
Unfortunately, we confirm previous findings \cite{hils2020measuring} that many CMP websites do not properly implement these standardized APIs, which makes it difficult to make use of them.
Moreover, CMPs are almost non-existent for very popular websites, which again leads to a lack of standardization potential for websites most visited by users.
Additionally, many cookie banners make it purposefully difficult for people to reject all cookies \cite{soe2020circumvention}.
As a prominent example, Google has been fined 150 million € for not providing users a choice to reject all cookies and was consequently forced to update their cookie banner \cite{google_banner}.
All these factors hinder effective banner automation and it is unlikely that the situation will improve without a joint push by browser developers, advertising companies, and lawmakers.

\noindent\textbf{Looking ahead:}
In order to improve user privacy, browser vendors have recently started to block third-party cookies at various degrees.
Mozilla introduced ``Enhanced Tracking Protection'' in 2019 \cite{mozilla_etp} and is now moving towards completely isolated cookie stores per website \cite{mozilla_tcp}.
Apple has introduced by-default TP cookie blocking in 2020 \cite{apple_tp,apple_tp2}.
Google has long touted its desire to get rid of TP cookies and proposed a myriad of different possible replacements \cite{google_ps,google_ps2, google_ps3,google_floc,google_topics}.
Getting rid of TP cookies is likely not the end of user tracking, as different techniques such as Local Storage, IndexedDB, Web SQL, or browser fingerprinting \cite{laperdrix2020browser} can easily replace TP cookie functionalities \cite{google_ending}.
Finally, privacy regulations such as GDPR are not specifically limited to cookies, but require informed consent for any shared user data, irrespective of the used technology.
Cookie banners will therefore likely remain a prominent sight in the future, even if the underlying technology might change.

\noindent\textbf{Limitations:}
Even though we cover a wide range of factors in our work, there are natural limitations to our approach.
First, since our banner detection approach leverages words from 12 languages,
we might not be able to detect banners on websites using other languages.
Second, we use \openwpm which uses the Firefox browser to access websites.
Websites could exhibit different cookie behavior when being accessed from a different browser, such as Chrome or Safari.
Third, we solely focus on HTTPS when accessing websites.
Since many browsers use an HTTPS-first approach and most websites do support HTTPS \cite{felt2017measuring}, we think this focus is warranted.
Websites can also be accessed via QUIC, which is not yet widely deployed \cite{zirngibl2021s}, and we thus do not consider it in our study.
Fourth, to classify third-party cookies as tracking cookies, we rely on tracking cookie lists.
In order to limit false positive tracking classifications, we use the conservative approach by Götze et al. \cite{gotze2022measuring}.
Therefore, our identified tracking cookies serve as a \textit{lower bound}.
\df{Fifth, to obtain the mobile version of the websites, we modify the \openwpm user agent and screen size (see \Cref{sec:mobilevsdesktop}). Although for most websites, we see the mobile version, for some websites these simple changes are not enough to load the mobile version \cite{yang2020comparative}.}
\section{Related Work}

To regulate the use of cookies, various data protection laws such as the GDPR \cite{GDPR} in the EU or CCPA \cite{CCPA} in California have been enacted in the last years.
A large body of previous work attempts to quantify the efficacy of such laws.
Dabrowski \etal \cite{dabrowski2019measuring} reported less persistent cookie usage for EU users in comparison to US users with Alexa top-100k websites as targets. On the contrary, Sanchez \etal \cite{sanchez2019can} claimed that the US appears to approach cookie regulations similar to the EU.
We do, however, observe a lower number of TP cookies in the EU when compared to non-EU VPs (see \Cref{sec:location}).

Furthermore, to check whether website publishers adhere to the EU cookie laws, Trevisan \etal \cite{trevisan2019} developed the tool ``CookieCheck'' \cite{cookiecheck}.
They reported that half of the websites they tested ($\approx 35k$) from an Italian VP, violate the law \ie they install profiling cookies\footnote{These are cookies that are managed by Web trackers to identify users and are clearly subject to explicit consent according to the GDPR.} before the user’s consent.
In contrast, we observe that in the no-interaction mode, ``only'' about 30\% of websites set tracking cookies at our EU VPs.
This might indicate that website publishers are adhering more to privacy laws over time.

While studying tracking, Iordanou \etal \cite{iordanou2018tracing} identified the geographic locations of the tracking servers.
They found that around $90\%$ of the tracking flows originating in the EU terminate at tracking servers hosted within the EU itself.
Additionally, there are multiple measurement studies that highlight how trackers use cookies for user profiling \cite{gonzalez2017cookie, falahrastegar2014rise, englehardt2016online,li2015trackadvisor, schelter2016tracking, Lerner2016, cahn2016empirical}.
As an example, Englehardt \etal \cite{englehardt2015cookies} demonstrated that adversaries could reconstruct up to $73\%$ of a user's browsing history using only the collected cookies.

Linden \etal \cite{linden2020privacy} took a different direction; they conducted a longitudinal study to assess privacy policies adopted by website publishers before and after GDPR went into effect.
They reported that GDPR has a positive impact on privacy policies.
Post-GDPR, not only the visual (and textual) representation of policies have improved, but the coverage of important topics \eg data retention, has also increased.
Degeling \etal \cite{degeling2019we} also made similar observations \ie after GDPR, many websites have added and updated their privacy policies and now show cookie banners to the users.
Sørensen \etal \cite{sorensen2019before}, rather than analyzing the privacy policies, found that after the introduction of GDPR, the number of third parties on EU websites has declined.
They noted, however, that it cannot be concluded with certainty that this decline is solely due to GDPR.
Kretschmer \etal \cite{kretschmer2021cookie} conducted a comprehensive survey of the existing research ($>70$ research papers), describing the legal as well as technical aspects of GDPR. They report that the enactment of GDPR has resulted in a decline in third-party tracking, increase in cookie banners, and privacy policies in the EU region. 

Santos \etal \cite{Santos2021Banners} studied cookie banners to analyze how clearly they  explain privacy policies. They manually analyzed $400$ cookie banners on English language websites that are popular in the EU. They report that $61\%$ of banners used vague language and violated the specificity purpose.
Utz \etal \cite{utz2019informed} rather than only focusing on the text of the banners, also studied other factors that could influence user consent decisions (\eg positioning of the banners on the website). The authors partnered with an e-commerce website in Germany and reported that changing the position of the banner or the text has a significant impact on the users' consent decisions. For instance, if the banner is shown in in the lower left part of the screen, users are more likely
to interact with it.

More recently, Chen \etal \cite{Chen2021} conducted a user survey of Californian consumers to study, to analyze how well they understand privacy policies of popular websites. They reported a significant variance in how websites interpret CCPA. Thus, privacy policy disclosures (mandated by CCPA) seem ambiguous to end-users. To this end, Connor \etal \cite{Connor2021} performed a study to specifically analyze how websites implement ``right to opt-out of the sale of users' personal information''. They observed that websites implement this mandate in ambiguous ways, which deters the users' motivation to opt-out.

Finally, other research specifically analyzes cookie banners themselves \eg how clearly they specify privacy policies \cite{Santos2021Banners} or the impact of banner location on user consent \cite{utz2019informed}. Jha \etal's \cite{jha2021internet} work is closest to our research. Similar to our work, the authors also attempted to interact with the banners in an automated manner to observe differences in cookies. However, their tool only accepts the privacy policies (of the banner), whereas our tool \tool has the capability to accept as well as reject a banner's consent.

\section{Conclusion}

In this paper, we performed a multi-perspective analysis of Web cookies.
We developed \tool to automatically detect, \df{accept, and reject cookie banners with an accuracy of 99\%, 97\%, and 87\%, respectively}.
Then we ran measurements from 8 geographic locations on 5 continents and identified substantial differences between these vantage points.
We found 56\% more banners on websites when visited from an EU vantage point.
Moreover, we quantified the effect of banner interaction: websites sent $5.5\times$ more third-party cookies on average after clicking ``accept''.
Accordingly, we observed a similar trend for tracking cookies as well.
Finally, we also identified differences in cookies depending on the visited page on a website (inner vs. landing) and the client platform (desktop vs. mobile).

\bibliographystyle{splncs04}
\bibliography{References}

\appendix

\section{HTML Elements Not Part of Cookie Banners}
\label{app:Element_property}

While detecting banners, if an element has words from our corpus (see \Cref{subsec:banner_detection}), and one of the following properties applies, we simply discard the element and move to the next one:
\one If the element is set as \emph{invisible}, the banner is not visible to users, and they can therefore not interact with it.
\two An element with a \emph{negative z-index} is behind some other objects on the page.
Thus it cannot contain a banner as the banner should be on top of every object in order to be visible by the user.
\three The banner should be within the user's visible area of a web page.
An element \emph{outside the viewport} cannot contain the banner.
\four The GUI part of a banner is generally not implemented using \emph{JavaScript}.
Thus even if it contains cookie-related words, we simply discard them.

We use additional heuristics \eg if the cookie-related words are present in a \texttt{table} element, we simply ignore them as well. We \df{make our code publicly available \cite{bannerclick}}, along with detailed information about these additional heuristics.

\section{Corpus of Words Used for Banner Interaction}
\label{app:corpus}

To create the corpus of the ``accept'', ``reject'', and ``settings'' words, we access the Tranco top-10K websites and detect the banners on them.
We proceeded with those Tranco websites, for which we successfully detect the banner.
Next, we identify the language of each of these websites using Google's \texttt{cld3} library \cite{cld3}.\footnote{\texttt{cld3} at its core uses neural networks to detect the language of any given document. We manually select 20 websites belonging to 10 different languages (\ie two websites for each language). We identify the language of these websites using \texttt{cld3} library and find it to be 100\% accurate.}
We observe that 4215 of these websites are in 12 languages; English alone is the language of more than 77\% of those.

\begin{wrapfigure}{l}{.5\textwidth}
\centering
\includegraphics[width=.5\textwidth]{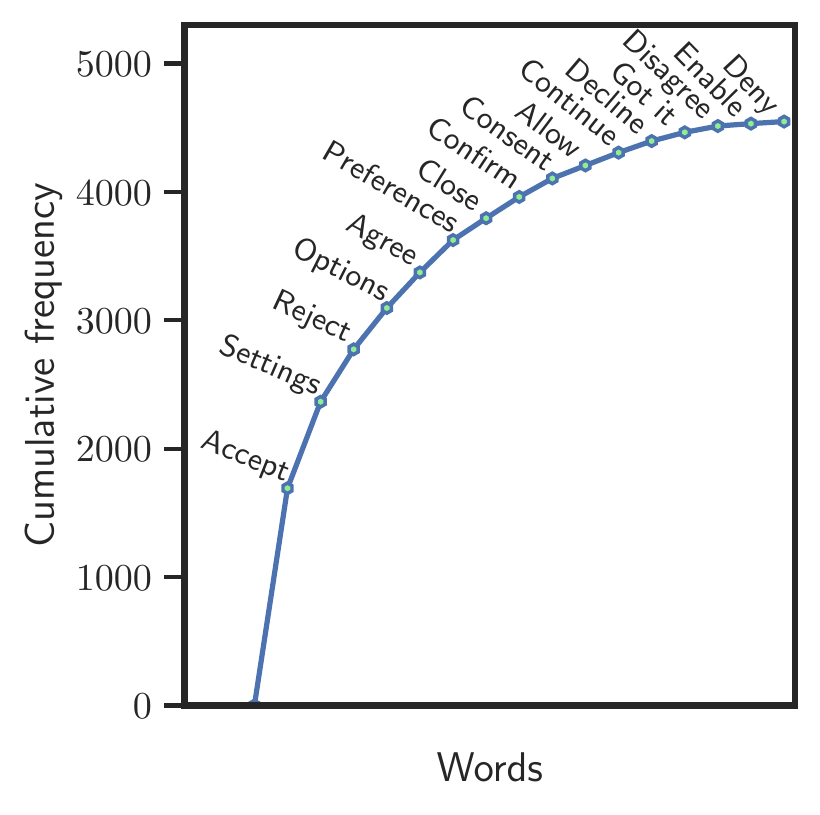}
\caption{Select English words appearing at differing frequencies inside the buttons of cookie banners.}
\label{fig:EN-words-interaction}
\end{wrapfigure}

To detect commonly used words in a given language, we adopt a simple approach.
For example, we select all banners in the English language, identify the \texttt{<buttons>} and their associated words in the banner, and count the frequency of such words.
We separate out the words that individually appear in at least $1\%$ of the banners. 
\Cref{fig:EN-words-interaction} shows examples for such words.
Examples for such words are ``Accept'', ``Settings'', ``Reject'', ``Options'', or ``Agree''.

For non-English languages (\eg German), we repeat the same process, but we additionally translate each of these words to English.
We then manually check if they are semantically similar to any one of the following three categories: accept, reject, or settings.
If the tested word is closer to any of these, we append the word to the appropriate category.
We repeat the same process for each of the 11 non-English languages.
At the end, we have %
172 words in 12 different languages belonging to the three different categories.

\clearpage

\section{Comparison With Priv-Accept Web Crawler}

Recently Jha \etal \cite{jha2021internet} proposed the tool Priv-Accept \cite{priv-accept}, which automatically attempts to ``accept'' privacy policies mentioned in a banner. They create a corpus of ``accept'' related words and compare them with the words present in the DOM of the website. If Priv-Accept finds the accept button, clicks it and compares the website behaviour before and after the click (\eg page load time).

We compare \tool with Priv-Accept. First, Priv-Accept is unable to identify and click reject buttons. Second, unlike \tool, Priv-Accept does not detect the banners but instead inspects the complete DOM for accept-related words and, on a successful match, attempts to click the element containing the word. As a result, it can encounter multiple failures before actually clicking the desired accept button on the banner. On the contrary, \tool first detects the banner and searches for words contained within the banner. 
Third, \tool can click on accept related elements in 12 popular languages whereas, Priv-Accept only searches for English words.
There are other differences, \eg \tool looks for banners within the iframes, but Priv-Accept ignores iframes.

We compare both tools on the Tranco top-1k websites.
With Priv-Accept, we can click accept on 451 websites, whereas with \tool, the number is 430. Websites where Priv-Accept could click accept but not \tool are 66, and vice-versa 59 websites.
The vast majority of the former set are websites that do not show an explicit accept option.
These are not considered to be explicit accepts by \tool, however Priv-Accept considers them.
Additionally, Priv-Accept also clicks on the incorrect accept button for \df{11 websites.}
The latter group contains websites where Priv-Accept is unable to identify the correct button, \tool detects banners in iframes, or the website is in a non-English language.

\end{document}